# A revised de Broglie relation in discrete space-time

Rui Qi

Institute of Electronics, Chinese Academy of Sciences

17 Zhongguancun Rd., Beijing, China

E-mail: rg@mail.ie.ac.cn

We introduce a revised de Broglie relation in discrete space-time, and analyze some possible inferences of the relation.

## Introduction

As we know, present theories including quantum mechanics and general relativity are based on the assumption of continuous space-time. It is implicitly assumed that space-time is continuous and infinitely divisible in these theories. But whether space-time is continuous or not is an unsolved problem. In fact, the proper combination of quantum mechanics and general relativity has implied the existence of discrete space-time, in which the minimum space and time unit is respectively Planck length and Planck time[1-2]. In this paper, we will assume space-time is discrete, and further analyze a revised de Broglie relation and its inferences.

## Why discrete space-time?

It can prove that when considering both quantum mechanics and general relativity, the minimum measurable time and space size will no longer infinitesimal, but finite Planck time and Planck length. Here we will give a well-known operational demonstration. Consider a measurement of the length between points A and B. At point A place a clock with mass $m$ and size $a$ to register time, at point B place a reflection mirror. When $t = 0$ a photon signal is sent from A to B, at point B it is reflected by the mirror and returns to point A. The clock registers the return time. For the classical situation the measured length will be $L = \frac{1}{2}ct$, but when considering quantum mechanics and general relativity, the existence of the clock introduces two kinds of uncertainties to the measured length. The uncertainty resulting from quantum mechanics is: $\delta L_{QM} \geq (\frac{\hbar L}{mc})^{1/2}$, the uncertainty resulting from general relativity is: $\delta L_{GR} \geq \frac{Gm}{c^2}$, then the total uncertainty is: $dL = dL_{QM} + dL_{GR} \geq (L \cdot L_p^2)^{1/3}$, where $L_P = (\frac{G\hbar}{c^3})^{1/2}$, is Planck length. Thus we conclude that the minimum measurable length is Planck length $L_P$. In a similar way, we can also work out the minimum measurable time, it is just Planck time $T_P = (\frac{G\hbar}{c^5})^{1/2}$.

## A revised de Broglie relation and its inferences

As we know, the usual de Broglie relation is applicable in continuous space-time. But in discrete space-time, it will be essentially revised due to the existence of minimum space-time unit. The reason lies in the following fact, i.e. as to the particles with very large energy, say larger than Planck energy, the corresponding de Broglie wavelength will be smaller than the minimum space size---Planck length according to the usual de Broglie relation. This contradicts the assumption of discrete space-time.

In order to satisfy the requirement of space-time discreteness, the simplest revised de Broglie relation can be written as follows[*]:

$$\lambda = \frac{h}{p} + \frac{1}{4}\frac{L_p^2 p}{h} \quad \text{---(1)}$$

$$T = \frac{h}{E} + \frac{1}{4}\frac{T_p^2 E}{h} \quad \text{---(2)}$$

It can be seen that the direct inference of these relations is $\lambda \geq L_p$ and $T \geq T_p$, which is consistent with the assumption of discrete space-time.

Before we can deduce the possible results of the revised de Broglie relation, another relation between energy and momentum is needed. We assume $p = mv_g$ and $E = mc^2$. This

---

[*] The precise form of such revised de Broglie relation may be $\lambda = \frac{h}{p} e^{\frac{L_p^2 p^2}{4h}}$ and $T = \frac{h}{E} e^{\frac{T_p^2 E^2}{4h}}$. This is equivalent to the energy-momentum transformation $p' = p e^{-\frac{L_p^2 p^2}{4h}}$ and $E' = E e^{-\frac{T_p^2 E^2}{4h}}$ or $E' = E e^{-\frac{E^2}{4E_p^2}}$, where $p', E'$ satisfy the usual relativistic relation, and their value space is respectively $[-\frac{h}{L_p}, \frac{h}{L_p}]$ and $[-\frac{h}{T_p}, \frac{h}{T_p}]$. The corresponding revised Schroedinger equation is

$i\hbar e^{kT_p^2 \frac{\partial^2}{\partial t^2}} \frac{\partial}{\partial t}\psi(x,t) = -\frac{\hbar^2}{2m} e^{2kL_p^2 \frac{\partial^2}{\partial x^2}} \frac{\partial^2}{\partial x^2}\psi(x,t) + V(x)\psi(x,t)$, where $k = \frac{1}{16\pi^2}$.

assumption is consistent with the to-be-deduced dispersion relation $E = pc$ for photons. It can prove that if assuming the relation $p = mv_g$ for any particles and the relation $E = pc$ for photons, then the relation $E = mc^2$ can be deduced using the conservation principle of energy-momentum. The deduction is irrelevant to the concrete form of the mass $m$. It should be denoted that the relation $v_g = \partial E/\partial p$ is no longer valid, since it results from the basic relation $v_g = \partial \mathbf{w}/\partial k$, which will be also revised on the condition of the revised de Broglie relation.

The first result is the revised dispersion relation, it is:

$$E^2 - p^2c^2 = m_0^2 c^4 + \frac{3}{8}\frac{E^2 + p^2c^2}{E_p^2}(E^2 - p^2c^2) \quad \text{---(3)}$$

or

$$E^2 - p^2c^2 = m_0^2 c^4 (1 + \frac{3}{8}\frac{E^2 + p^2c^2}{E_p^2}) \quad \text{---(4)}$$

The nonrelativistic approximate relation is:

$$E = \frac{p^2}{2m}(1 - \frac{L_p^2 p^2}{2h^2}). \quad \text{---(5)}$$

As an example, as to the ideal infinite square potential well, the revised energy level formula is $E_n' = E_n(1 + \frac{T_p^2 E_n^2}{4h})$, where $E_n = \frac{n^2 h^2}{8mL^2}$. As we can see, in such discrete space-time, the usual dispersion relation $E = pc$ for photons still holds, and the light speed is still irrelevant to its wavelength.

The relativistic mass formula turns to be:

$$m = [\mathbf{g} + \frac{3E_0^2}{16E_p^2}\mathbf{g}^3]m_0. \quad \text{---(6)}$$

This indicates that the relativistic space-time transformation will be also revised. It can be seen that whatever the revised space-time transformation is the revised de Broglie relation will always satisfy the requirement of discrete space-time, i.e. the minimum space and time unit will be Planck length and Planck time in any inertial frame.

Furthermore, we can work out the revised uncertainty relation using the relation

$\Delta p^2 = \overline{p^2} - \overline{p}^2$, it is:

$$\Delta x \cdot \Delta p \geq h + \frac{L_p^2 \Delta p^2}{4h} \quad \text{---(7)}$$

One of the direct results of the uncertainty relation is $\Delta x \geq L_p$, this is consistent with the assumption of discrete space-time. Besides, the commutative relation will be also revised, it is:

$$[x, p] = ih \cdot (1 + \frac{L_p^2 p^2}{h^2}) \quad \text{---(8)}$$

This seems to be equivalent to the transformation of Planck constant $h \to h \cdot [1 + \frac{L_p^2 \overline{p}^2}{h^2}]$.

## Another two possibilities

Lastly, we consider the other two possibilities of discrete space-time assumption.

(1). We assume that space is discrete, but time is continuous, i.e. there only exists the minimum space unit. Then the revised de Broglie relation turns to be $l = \frac{h}{p} + \frac{1}{4} L_p^2 p$ and $T = \frac{h}{E}$. This directly results in the relation $l \geq L_p$ and $T \geq 0$, which satisfies the assumption of discrete space. In such space-time, the light speed will be related to its wavelength.

The revised dispersion relation for photons is $E^2 = p^2 c^2 [1 - \frac{3L_p^2 p^2}{8h^2}]$ [†]. The relation between light speed and wavelength is $v_g = c[1 + \frac{3L_p^2 p^2}{16h^2}]$. The revised uncertainty relation is still $\Delta x \cdot \Delta p \geq h + \frac{L_p^2 \Delta p^2}{4h}$.

(2). We assume that time is discrete, but space is continuous, i.e. there only exists the minimum time unit. Then the revised de Broglie relation turns to be $l = \frac{h}{p}$ and $T = \frac{h}{E} + \frac{1}{4} \frac{T_p^2 E}{h}$. This directly results in the relation $l \geq 0$ and $T \geq T_p$, which satisfies the assumption of discrete space. In such space-time, the light speed will be also related to its wavelength.

---

[†] The general dispersion relation is: $E^2 = m_0^2 c^4 + p^2 c^2 [1 - 3L_p^2 p^2 / 8h^2]$.

The revised dispersion relation for photons is $E^2 = p^2c^2[1+\frac{3L_p^2 p^2}{8h^2}]$. The relation between light speed and wavelength is $v_g = c[1-\frac{3L_p^2 p^2}{16h^2}]$. The revised uncertainty relation is still $\Delta x \cdot \Delta p \geq h + \frac{L_p^2 \Delta p^2}{4h}$.

## Discussions

As we know, the macroscopic object can easily possess the energy larger than Planck energy, but it is evident that the revised dispersion relation doesn't hold true for them. Then why? i.e. why the revised de Broglie relation is only applicable for the microscopic particles, but not applicable for the macroscopic objects? Some may think that the reason should be relevant to the structure of matter. Here we will argue that the reason lies in the environmental decoherence and dynamical collapse of wave function.

As to the microscopic particle, the effect of environmental decoherence is generally very weak, and the dynamical collapse time of its wave function is very very long, or we can say, its wave function doesn't collapse. Thus the motion of microscopic particle is still quantum motion which possesses wave property, and the revised de Broglie relation is applicable for the microscopic particles.

But as to the macroscopic object, the effect of environmental decoherence is very strong, and the dynamical collapse time of its wave function is very very short. Then long before its energy reach the Planck energy during its formation its wave function will have collapsed many times, and its motion will be not quantum motion, but approximately continuous motion, which satisfies the classical motion equation. This indicates that the macroscopic object will basically lose its wave property, so the revised de Broglie relation and dispersion relation describing the wave property will be not applicable to it. Furthermore, according to the Enrenfest theorem, as to the macroscopic object with energy very smaller than Planck energy, the dispersion relation should be the revised dispersion relation in which the revised term can be omitted, namely

$E^2 - p^2c^2 = m_0^2 c^4$. Then when the energy of macroscopic object turns to be larger and larger, its motion is still continuous motion without wave property, so the dispersion relation will be also the same as the above one. It should be denoted that if the macroscopic object is well isolated from the environment, it motion will also satisfy the revised dispersion relation as the microscopic particles.

## Conclusions

In this paper, a revised de Broglie relation in discrete space-time is introduced, and its possible inferences .are analyzed. We denote that the revised de Broglie relation does not hold true for macroscopic object due to the environmental decoherence and dynamical collapse of wave function.